\documentclass[12pt]{article}
\input epsf.sty
\def\ZZZ{{\hbox{ Z\kern-1.6mm Z}}}
\def\beq{\begin{equation}}
\def\eeq{\end{equation}}
\def\bea{\begin{eqnarray}}
\def\eea{\end{eqnarray}}
\def\R{\rangle}
\def\L{\langle}
\def\lt{\left}
\def\rt{\right}
\def\l{\lambda}
\def\lp{\lambda_{+}}
\def\lm{\lambda_{-}}
\def\del{\partial}

\def\Ip{{\cal I}^+}

\def\pin{\phi^{in}}
\def\pins{\phi^{in *}}
\def\po{\phi^{out}}
\def\pos{\phi^{out *}}

\def\qt{\tilde q}

\def\a{\alpha}
\def\b{\beta}

\def\wp{w^{\prime}}
\def\wb{\bar w}

\def\ai{a^{in}}
\def\ao{a^{out}}

\def\ati{\tilde a^{in}}
\def\ato{\tilde a^{out}}

\def\Ai{A^{in}}
\def\Ao{A^{out}}

\def\Aod{A^{out \dagger}}

\def\abi{\bar a^{in}}
\def\abo{\bar a^{out}}
\def\abid{\bar a^{in \dagger}}
\def\abod{\bar a^{out \dagger}}

\def\gs{g_s}

\def\fl{f_{\lambda}}
\def\flp{f^{\prime}_{\lambda}}

\def\gl{g_{\lambda}}

\def\0{0+}

\def\Il{I_{\lambda}}
\def\g{{\bf g}}

\def\Do{D^{out}}
\def\du{\delta u}
\def\dv{\delta v}

\def\hl{h_{\lambda}}
\def\hlp{h_{\lambda}^{\prime}}
\def\hlpp{h_{\lambda}^{\prime \prime}}
\def\th{\tilde h}
\def\E{{\cal E}}
\def\P{{\cal P}}

\def\Brho{B^{(\rho)}}
\def\Bmu{B^{(\mu)}}
\def\Hl{H_{\lambda}}
\def\Irho{I^{(\rho)}}
\def\Imu{I^{(\mu)}}
\def\G{\Gamma}

\newcommand{\sectiono}[1]{\section{#1}\setcounter{equation}{0}}

\begin{document}

{}~
{}~
\hfill\vbox{\hbox{UK-04/12} \hbox{hep-th/0406029}}\break

\vskip .6cm

\centerline{\Large \bf
On The Problem of Particle Production in $c=1$ Matrix Model}

\medskip

\vspace*{4.0ex}

\centerline{\large \rm Partha Mukhopadhyay }

\vspace*{4.0ex}

\centerline{\large \it  Department of Physics and Astronomy}

\centerline{\large \it  University of Kentucky, Lexington, KY-40506, U.S.A.}

\medskip

\centerline{E-mail: partha@pa.uky.edu}

\vspace*{5.0ex}

\centerline{\bf Abstract} \bigskip

We reconsider and analyze in detail the problem of particle 
production in the time dependent background of $c=1$ matrix model 
where the Fermi sea drains away at late time. In addition to the
moving mirror method, which has already been discussed in 
hep-th/0403169 and hep-th/0403275, we describe yet another method of
computing the Bogolubov coefficients which gives the same result. 
We emphasize that these Bogolubov coefficients are approximately 
correct for small value of the deformation parameter.

We also study the time evolution of the collective field theory
stress-tensor with a special point-splitting regularization. Our
computations go beyond the approximation of the previous treatments and
are valid at large coordinate distances from the 
boundary at a finite time and up-to a finite coordinate distance from 
the boundary at late time. In this region of validity our
regularization produces a certain singular term that is precisely
canceled by the collective field theory counter term in the present
background. The energy and momentum densities fall off exponentially
at large distance from the boundary to the values corresponding to the 
static background. This clearly shows that the radiated energy reaches
the asymptotic region signaling the space-time decay.

\vfill \eject

\baselineskip=16pt

\tableofcontents

\section{Introduction and Summary} 
\label{intro}
Two dimensional bosonic and type 0B string theories have 
non-perturbative dual description in terms of the $c=1$ matrix 
model\footnote{The subject is well developed. Various reviews and 
some original papers can be found in 
refs.\cite{polchinski-book, klebanov91, ginsparg93, jevicki93, 
c=1, takayanagi03, douglas03}.}. 
Although the CFT description is quite complicated 
\cite{dorn92, dorn94, zamolodchikov95, teschner01}
the matrix model description is simple. In the singlet sector 
this reduces to the quantum mechanics of an infinite (but fixed) 
number of non-relativistic free fermions in an inverted harmonic oscillator
potential. Recently there has been a crucial progress in this subject 
through the understanding of the unstable D0 brane \cite{Dbrane} and its 
decay on the matrix model 
side \cite{martinec03, alexandrov03, mcgreevy03,
klebanov03}\footnote{See \cite{sen03, recent} for relevant recent works.}. 
Here also the matrix model description \cite{klebanov03} of this decay
turns out to be quite simpler than the BCFT description \cite{sen02}.
It is 
this simplicity of the matrix model description and the fact that it 
provides us with a non-perturbative definition of the theory enable us
to probe various issues that would have been difficult to address
otherwise (see, for example, \cite{sen03}). Therefore it is important 
to study various other
backgrounds, in particular the time dependent backgrounds
\cite{minic92, moore92, alexandrov02, karczmarek03, karczmarek04, 
das04, karczmarek042, ernebjerg04}, in the matrix model itself to help 
understanding the corresponding world-sheet theories. This might
provide us with clues on how to deal with time dependent backgrounds
in string theory in general.

Recently in \cite{karczmarek03}, some interesting fermionic 
configurations in the matrix model were viewed as two 
dimensional cosmological solutions. A particular 
class of time dependent solutions 
with two parameters were discussed where the envelop of the Fermi sea, 
having the same structure as the static one, has an overall motion on the 
single particle phase space. The motion is such that the Fermi sea
floods in and subsequently drains away thus
describing creation and the subsequent annihilation of the universe.
They have been interpreted to be arising from the closed string
tachyon condensation \cite{karczmarek03, karczmarek04,
  karczmarek042}. The corresponding world-sheet deformation was
also suggested. Switching off one of the two
parameters one reaches two different classes of one parameter
solutions where the Fermi sea merges with the static one in either the
past or the future asymptote.
In this case the deformation parameter does not really parameterize
inequivalent solutions as the change in this parameter can be absorbed
by time translation\footnote{All these solutions very much look like
the bulk analogs of the boundary rolling tachyon solutions discussed
in \cite{sen02}. The one parameter deformations are analogous to the
situations where the open string tachyon starts from at early time or
reaches at late time the top of the potential. In the open string
context they are interpreted to be either creation or destruction
of the D-brane whereas in the closed string context they correspond
to creation or annihilation of the space-time itself. One crucial
difference between the two cases is that open string solutions arise
from a perturbative instability whereas the closed string ones should
be considered as different backgrounds altogether \cite{karczmarek03}
because the
deformations are by non-normalizable modes (there is no perturbative 
instability).}.
The solution, hereafter called the ``draining Fermi sea'', in which the
Fermi sea starts from a configuration arbitrarily close to the static
one at early time and drains away at late time has been further
studied in \cite{karczmarek04, das04} where 
the problem of cosmological particle production has been addressed. In
particular in \cite{das04} the Bogolubov coefficients for the particle
production have been computed explicitly. An approximate analysis of
the energy-momentum tensor was also performed and it was argued that
the contribution to the energy coming from particle production and the
time evolution of the initial vacuum energy cancel so that the net
energy remains the same.
   
In this paper we study the same questions in more detail.
In the static background the
theory is free far away from the tachyon wall. In the draining
background the tachyon wall moves in such a way that an observer
initially sitting in the free region gradually enters the strongly
coupled region where the notion of the massless particle is lost and 
the theory becomes more complicated. Nevertheless there is a time 
span during which the
observer stays in the free region and can observe particle
production. We attempt to quantify this observation during this
particular span of time. Although the deformation parameter $\l$
corresponding to this background can be changed by time translation we
fix the origin of time and treat $\l$ as an actual parameter. In this
time-frame the condition for the
observer not to enter the strongly coupled region during the whole
process of observing particle production turns out to be 
$\ln1/2\l >> 0$. We shall see that this condition will be essential
for computing approximate Bogolubov coefficients for the particle
production. 

In the first analysis of the system we study the computation
of the Bogolubov coefficients in two different methods namely, the 
``scattering method" and the ``moving mirror method''. The first
method is directly related to the wall-scattering\footnote{ 
See, for various methods of computing the scattering amplitudes,
\cite{gross91, moore922, demeterfi, sengupta91, difrancesco91, 
polchinski91, moore92, polchinski942}.} in the static background and 
uses crucially 
the Seiberg 
bound \cite{seiberg90, polchinski902} on the spectrum of
primary operators in the CFT 
\cite{dorn92, dorn94, zamolodchikov95, teschner01}. The space-time version
\cite{ginsparg93} of this bound is that the in-states of this
particular scattering problem contains only the right-moving
excitations which move toward the wall. Similarly the out-states
contain only the left moving excitations. Application of the same
bound in the time dependent case implies that the in and 
out-vacua support only the right and left-moving excitations 
respectively. In the static case the result of the classical
wall-scattering is 
given by a ``scattering equation" \cite{polchinski91} where a
right-moving 
oscillator is
expressed in terms of the left-moving oscillators in the form of a
power series, the leading term being linear and of order $\gs^0$. 
Therefore computing the scattering equation up-to the leading order in
the time dependent case should give us the Bogolubov coefficients. 
Although this scattering equation is classical, non-trivial Bogolubov 
coefficients, which indicate quantum phenomenon like particle
production, can be obtained with the following additional input. The 
motion of the moving Fermi sea on the phase-space can be undone by a 
time dependent canonical transformation which can be lifted to a 
coordinate transformation on the collective field 
\cite{karczmarek03, das04}\footnote{Following 
\cite{das04}, this can also be viewed as the 
action of a particular $W_{\infty}$-transformation \cite{Winfty} 
on the collective 
field.}. Implementing this coordinate transformation on the scattering
equation in the static background gives non-trivial mixing between the
positive and negative frequency modes. It turns out that this analysis 
gives approximate Bogolubov coefficients for small $\l$ such that 
$\ln1/2\l >> 0$.

In the moving mirror method, which has already been discussed in
\cite{karczmarek04, das04}, 
we set up the whole problem in the framework of the Das-Jevicki
collective filed theory \cite{das90}. We work at the linearized level 
of the equation of motion which corresponds to taking a formal $\gs \to 0$
limit\footnote{The effective coupling, which has $\gs$ as an overall
factor, increases gradually as we approach the wall. By choosing
$\gs$ arbitrarily small we can increase the region of validity of this 
approximation.}. In this
approximation the problem of particle production in the
moving Fermi sea backgrounds reduces to that of a moving mirror problem
with reflecting boundary condition. This subject has been studied
quite extensively in flat-space 
\cite{dewitt75, davies76, davies77, birrell}. 
In our case, it turns out that the
metric is time dependent and therefore, as expected, introduces problems in
defining the natural modes that should correspond to
particles. In flat space examples of moving mirror one uses a certain
argument of geometrical optics to construct the natural in and
out-modes. We generalize this construction to the present case
at hand.

The class of moving Fermi sea backgrounds are such that asymptotically 
the mirror approaches the velocity of light. 
In the draining Fermi sea background, which we concentrate on, 
this happens at the future asymptote where the mirror moves toward 
the observer\footnote{Examples of mirror-trajectories approaching a 
null line has already been discussed in the literature
\cite{davies76, davies77} with the 
exception that mirrors moving away from the observer are usually
considered. One of the known examples reproduces Hawking radiation 
\cite{hawking75}.}. 
In a standard moving mirror example \cite{dewitt75} one assumes that
at far future the 
mirror approaches a constant velocity less than that of light so that the 
motion can be undone by a Lorentz transformation. In case of a mirror
approaching the velocity of light toward the observer the whole
portion of the future null asymptote is not available for defining the
modes. In fact the natural out-modes that are constructed
following the method of geometrical optics form an over-complete set in
that region which destroys the orthonormality of these modes. It turns
out that for the particular mirror-trajectory involved in our example 
the upper bound on the future null infinity is given by $\ln 1/2\l$. 
Therefore taking $\ln1/2\l >>0$ we make these out-modes approximately
orthonormal. The Bogolubov coefficients computed using these out-modes
are therefore approximately correct for small $\l$. 
As expected from the fact that $\l$ acts as a
deformation parameter, both the above methods give trivial Bogolubov
coefficients in the limit $\l \to 0$. Therefore our results are
approximately correct for small but nonzero $\l$.

Next we turn to the analysis of the stress-tensor and go beyond the 
approximation made in \cite{das04} (see the last part of the
discussion in sec.\ref{s:conclude} for more details). 
Incorporating the effect of the non-flat metric
we explicitly compute the vacuum expectation 
value of the stress-tensor by expanding field in terms of 
the in-modes constructed in the moving mirror discussion. 
Doing the computation in a point-splitting
regularization method we encounter the usual vacuum ambiguity.
Das-Jevicki collective field theory \cite{das90} automatically comes
with a particular counter term which fixes this ambiguity
(see \cite{das04} for a recent discussion). This counter term was
first obtained in \cite{andric85} using the collective field method 
\cite{jevicki80} which is formally background independent. It was also
showed by Gross and Klebanov in their ``fermionic string field theory''
formulation \cite{gross912} that this counter term is equivalent to
the normal ordering directly obtained from the fermionic theory.  
Considering the static case first, we show that in the 
 computation 
using the mode expansion the known result is reproduced by a 
special type of point-splitting regularization method. We generalize
this method to the time dependent case and obtain results for both
the singular and finite parts of the stress-tensor components. 
We justify this regularization by showing that the counter term in 
the Hamiltonian required to cancel the singular part precisely agrees 
with the collective field theory counter term evaluated at the present
background. We should mention at this point that our computations are
actually done in a metric which is simpler than the actual one. This
gives rise to a particular region of validity only where we achieve
the above agreement. The region of validity of our analysis is (1) 
everywhere at early time, (2) large coordinate distances from the
boundary at a 
finite time and (3) up-to a finite coordinate distance from the
boundary at late time. The finite part of the energy density
approaches the value corresponding to the static background at early 
time but evolves into something else at late time giving rise to a 
nonzero radiated energy. We show that the energy and momentum
densities fall off exponentially to the values corresponding to the 
static background at large distance from the boundary. 

The rest of the paper is organized as follows.
We review some of the basic relevant points of both the world-sheet
theory and the matrix model in sec.\ref{s:review}.
The time dependent backgrounds discussed in \cite{karczmarek03} are
reviewed in sec.\ref{s:MFS}. 
The two methods of computing the Bogolubov coefficients are given in
subsections \ref{ss:scattering} and \ref{ss:moving-mirror}. 
Sec.\ref{s:stress-tensor} contains the stress-tensor analysis.
Finally, we conclude in sec.\ref{s:conclude} 
where we summarize the basic accomplishments of this paper and compare
our stress-tensor analysis with those preexisted 
\cite{karczmarek04, das04, karczmarek042} in the literature. 
The appendices contain some of the technical details. We discuss the 
relation between the Bogolubov and the stress-tensor analysis in 
appendix \ref{A:relation}.
\section{Review of Basic Facts}
\label{s:review}
Here we touch upon the basic features of the world-sheet and the
matrix model descriptions relevant for our discussion and spell out
the dictionary of the duality. 
\subsection{The World-Sheet Theory}
\label{ss:ws}
In addition to the usual time-like scalar $X^0$, the matter part of 
the world-sheet theory contains only one space-like scalar $X^1$ whose
action is given by a particular limit of the Liouville action 
\cite{dorn92, dorn94, zamolodchikov95, teschner01},
\bea
S_L &=& \frac{1}{2\pi} \int d^2z \lt ( 
\del X^1 \bar \del X_1  + 2 \pi \mu_0 e^{2 b X^1} \rt) ~,
\label{SL}
\eea
with a background charge $Q = b + 1/b$ and central charge $c_L=1+6 Q^2$. 
The space of normalizable states is given by the collection of the
conformal families corresponding to a one parameter family of
primaries given by\footnote{Note that this form of the vertex operator
is valid only at large negative $X^1$ where the 
interaction term in (\ref{SL}) is negligible.},
\bea
V_{Q+iw} = e^{(Q+iw)X^1}~, \quad \quad 0 \leq w < \infty ~,
\label{VQw}
\eea
with conformal dimension $(Q^2 + w^2)/4$. Notice that $w$ is
restricted to be positive, the so-called Seiberg
bound \cite{seiberg90, polchinski902}. In fact the
operators with negative values of $w$ are related to the above ones by
the following reflection equation,
\bea
V_{Q+iw} &=& R(w) V_{Q-iw}~, \cr
R(w) &=& - \mu^{-iw/b} 
\frac{\Gamma(1+iw/b)\Gamma(1+iwb)}{\Gamma(1-iw/b)\Gamma(1-iwb)}~, \cr
\mu &=& \pi \mu_0 \gamma(b^2)~, \quad \gamma(x) = 
\frac{\Gamma(x)}{\Gamma(1-x)}~.
\label{reflection}
\eea
The world-sheet theory $S_{X^1}$ relevant for string theory is
obtained (at $\alpha^{\prime} =1$) by taking the following limit,
\beq
b \to 1~, \quad \mu_0 \to \infty~, \quad \mu ~\hbox{fixed}~.
\label{limit}
\eeq
It turns out that it is only the combination $\mu /g_s$ which appears 
as a parameter in all the amplitudes so that there is only one
parameter of the string theory which we take to be $g_s$.
BRST analysis of the world-sheet theory shows that there is only one
space-time field-theoretic degree of freedom, namely the 
``massless tachyon'' given by the vertex operators,
\bea
V_w^{\pm} = c \bar c e^{-iwX^0} V_{2\pm iw} = 
c \bar c e^{-iw(X^0 \mp X^1)} e^{2X^1} ~, \quad w>0 ~.
\label{vertex}
\eea
The space-time interpretation of the reflection equation
(\ref{reflection}) or the Seiberg bound $w>0$ in the above equation
goes in the following way (see, for example \cite{ginsparg93}). The 
effective coupling 
$g_{eff}=\gs e^{2x^1}$ falls off to zero at large negative $x^1$ where
we have a free massless particle \cite{polchinski902}. Because of the 
Liouville interaction-wall in (\ref{SL}) any right-moving pulse created
in the free region necessarily results into left-moving pulses due to
the scattering from the wall.
Therefore in this scattering problem the 
in  and out-states always contain only the right ($V_w^+$) and 
left-movers ($V_w^-$) respectively. In the second quantized theory 
these states are created by harmonic oscillators $\abid_w$ and
$\abod_w$ respectively. We take the following normalization.
\bea
[\abi_w, \abid_{\wp}] = [\abo_w, \abod_{\wp}] = \delta(w-\wp)~.
\label{a-comm}
\eea
Then one concludes from (\ref{reflection}) with the proper limit 
(\ref{limit}) that to the leading order in $\gs$, $\abi_w$ is
proportional to $\abo_w$ where the proportionality constant is simply 
a $w$ dependent phase. 
\subsection{The Matrix Model}
\label{ss:MM}
In the singlet sector, the 
c=1 matrix model reduces to the quantum mechanics of an infinite (but
fixed) number of non-relativistic free fermions in an 
inverted harmonic oscillator potential given by $-x^2/2$. 
$\gs$ plays the role of the Planck's constant $\hbar$ of the fermion 
theory. The string theory Hamiltonian is given by the second quantized
Hamiltonian of these fermions with an additional factor of 
$1/\gs$ \cite{polchinski91}.  
The closed string background discussed in the previous section 
corresponds to a particular classical state where all the 
single particle phase-space trajectories below energy $-1/2$ are 
filled by fermions. For the bosonic case, which is under consideration
in this paper, only those trajectories which are on the left side
of the potential are filled so that the envelop of the (static) 
Fermi sea is given by,
\bea
(x+p) (x-p) &=& 1~, \quad x \leq -1~.
\label{static-env}
\eea
\begin{figure}
\begin{center}
\leavevmode
\hbox{%
\epsfxsize=2.5in
\epsfysize=2.5in
\epsffile{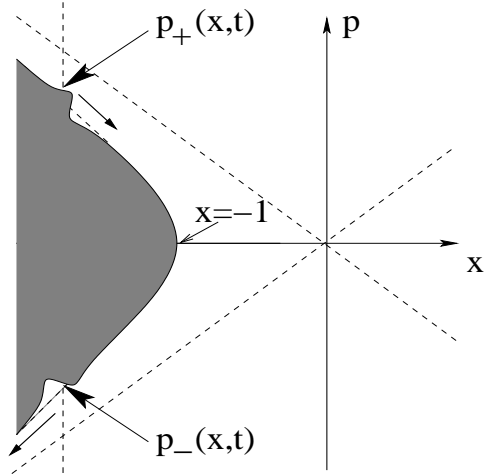}}
\caption{The ``static Fermi sea'' background which is dual to the
closed string background with linear dilaton and static tachyon wall. The
fluctuation of the upper and lower edges (which propagate along the
direction of the arrows) give rise to the massless tachyon field.}
\label{fig-SFS}
\end{center}
\end{figure}
Let $p_{\pm}(x,t)$ be the values of $p$ at the fluctuating upper and
lower edges (fig.\ref{fig-SFS}). Then the fluctuations 
$\eta_{\pm}(x,t)$ are defined as, 
\bea
p_{\pm}(x,t) = p_{0\pm}(x) + \eta_{\pm}(x,t)~,
\label{ppm}
\eea
where the background is given by,
\bea
p_{0\pm}(x)= \pm P_0(x)~, \quad P_0(x) = \sqrt{x^2 -1}~, \quad 
x\leq -1~.
\label{p0pm}
\eea
Using Polchinski's bosonization in \cite{polchinski91} we define
the fluctuations of the scalar field and the conjugate momentum,
denoted by $\bar \phi(x,t)$ and $\pi_{\bar \phi}(x,t)$ respectively,
\bea
\eta_{\pm}(x,t) = \sqrt{\pi} g_s \lt[- \pi_{\bar \phi}(x,t) \pm 
\del_x \bar \phi(x,t) \rt]~.
\label{etapm}
\eea
$x=-1$ acts as a boundary where $\bar \phi$ satisfies a Dirichlet
boundary condition\footnote{This can be derived from the constraint 
that the total number of fermions be fixed.}: $\bar \phi(x=-1, t) = 0$. 
The static Fermi sea background (\ref{p0pm}) is given, in terms of the
collective variables, as,
\bea
\del_x \bar \phi^{(S)}_0(x,t) = {P_0(x) \over \sqrt{\pi} g_s}~, \quad 
\pi_{\bar \phi^{(S)}_0}(x,t) =0 ~.
\label{coll-backgrd}
\eea
Defining the new coordinate $q$,
\bea
x = -\cosh q ~, \quad -\infty < q \leq 0~,
\label{x-q}
\eea
so that $P_0= |\sinh q|$, $P_0 \del_x = \del_q$, the linearized
equation of motion takes the form,
\bea
(\del_t^2 - \del_q^2) \phi = 0~,
\label{phi-eom}
\eea
where $\phi(q,t) = \bar \phi(x,t)$. It turns out that at $q<<0$ the
non-linear parts of the equation of motion reduce to 
an infinite series of interaction terms with 
$g_{eff}= g_s e^{2 \qt}~(\tilde q = q + \ln2)$ as an effective coupling.
In this region the fluctuation $\phi(q,t)$ can be mode expanded as,
\bea
\phi &=& \phi_{in}(u)  + \phi_{out}(v) ~, \cr
\phi_{in}(u) &=& \int_0^{\infty} {dw \over \sqrt{4 \pi w}} \lt[\ai_w
e^{-iwu} + h.c. \rt] ~,\cr
\phi_{out}(v) &=& \int_0^{\infty} {dw \over \sqrt{4 \pi w}} \lt[\ao_w
e^{-iwv} + h.c. \rt]~,
\label{mod-expand}
\eea
where $u=t-q$ and $v=t+q$ are the usual null coordinates. The
oscillators $\ai_w$ and $\ao_w$ have the same normalization as in 
(\ref{a-comm}). The scalar field $S(q,t)$ introduced in 
\cite{polchinski91}, is related to $\phi(q,t)$ at $q<<0$ in the following way:
$S(\qt,t) = \phi (q,t)$. Therefore the oscillators, say $\ati_w$ and 
$\ato_w$, of $S(q,t)$ are related to that of $\phi(q,t)$ as follows:
$\ai_w = e^{iw\ln 2} \ati_w$, $ \quad \ao_w = e^{-iw\ln 2} \ato_w $.
Using the result obtained in \cite{polchinski91}, we get the following
scattering equation,
\bea
\ai_w = e^{iw\ln 4} \ao_w + O(\gs)~.
\label{MM-scat0}
\eea
The oscillators of the scalar field directly coming from string theory
are related to the above oscillators through the leg-pole factors
\cite{gross91, polyakov91, difrancesco91}
with
the additional phase factor described above.
\bea
\abi_w = {\Gamma (iw) \over \Gamma (-iw)} e^{-iw\ln 2} \ai_w ~,\quad
\abo_w = {\Gamma (-iw) \over \Gamma (iw)} e^{i w \ln2} \ao_w~.
\label{leg-pole}
\eea

The above derivation of the collective field theory is classical.
Therefore in the process of quantization one encounters the usual 
vacuum ambiguity which has to be fixed by a normal ordering
prescription. In the string field theory of Gross and Klebanov 
\cite{gross912}, which was shown to be equivalent to that of Das
and Jevicki \cite{das90}, this automatically came from the normal 
ordering of the fermionic theory. 
In Das-Jevicki field theory this is incorporated by
a particular counter term first computed in \cite{andric85} using the
collective field method \cite{jevicki80}. In our notation this 
counter term is given by,
\bea
\Delta H = {\gs \over 2\sqrt{\pi}} \int_{-\infty}^{-1} dx ~\del_x \bar
\phi(x,t) \del_x \del_{x^{\prime}} \ln |x-x^{\prime}|_{x=x^{\prime}}~.
\label{Hct}
\eea
According to the collective field derivation \cite{andric85} this
counter term is background independent\footnote{I am thankful to
S. R. Das for discussion on this point.}. 
\section{The Moving Fermi Sea Backgrounds}
\label{s:MFS}
We shall now consider a particular class of two-parameter
time-dependent closed string backgrounds discussed in
\cite{karczmarek03}. On the matrix model side this is described by a 
moving Fermi sea whose envelop is given by,
\bea
(x+p+2 \lp e^t) (x-p+2\lm e^{-t}) &=& 1~, \quad x\leq x_m(t) ~,~~x_m(t)= 
-(1 + \lp e^t + \lm e^{-t})~.\cr &&
\label{moving-env}
\eea
The envelop of the Fermi surface is an hyperbola which moves like a
rigid body following the hyperbolic trajectory of its centre:
$x_0(t) = -(\lp e^t + \lm e^{-t})~, ~~ p_0(t)= -(\lp e^t - \lm e^{-t})$.
Just like as described in the previous section the fluctuation of this 
Fermi surface also gives rise to a ($1+1$)-dimensional scalar field
$\phi(t,q)$ which, at any given instant of time, behaves like a free 
massless field at a sufficiently large negative $q$. On the
phase-space this configuration can be transformed into the static sea 
by the following time-dependent canonical transformation\footnote{As
discussed in \cite{das04} this can also be viewed as a
particular $W_{\infty}$ transformation. },
\beq
X=x + \lp e^t + \lm e^{-t}~, \quad P = p + \lp e^t - \lm e^{-t}~,
\quad -\infty < X \leq -1~.
\label{can-transf}
\eeq
Following the usual method as described in subsection \ref{ss:MM} 
a $(1+1)$
dimensional scalar field theory can be constructed from the static 
configuration in the $XP$-plane. We shall, in this case, use upper 
case symbols for various objects discussed in the previous section.
We relate the space-time of this field theory with the original 
(matrix model) space-time $(x,t)$ by the following coordinate 
transformation,
\bea
X=x + \lp e^t + \lm e^{-t}~, \quad T=t~.
\label{coord-transf}
\eea
Following (\ref{x-q}) we define new coordinates $Q$ and $q$ as,
\bea
X &=& -\cosh Q~,\quad -\infty < Q \leq 0 ~, \cr
x &=& -\cosh q~,\quad -\infty < q \leq q_m(t) ~, \cr
q_m(t) &=& - \cosh^{-1}(1 + \lp e^t + \lm e^{-t})~,
\label{X-Q-x-q}
\eea
such that,
\beq
\cosh Q = \cosh q - (\lp e^t + \lm e^{-t})~.
\label{Q-q}
\eeq
At large negative $Q$ this reduces to 
$Q = q - \ln (1 - 2 \lp e^v - 2 \lm e^{-u})$ such that,
\bea
U = u + \ln (1 - 2\lp e^v - 2 \lm e^{-u})~, \quad V = v - \ln (1-2\lp
e^v -2\lm e^{-u})~,
\label{UV-uv}
\eea
where as usual we have defined: $U=T-Q,~ V=T+Q$.
The scattering equation for the oscillators $\Ai_w$ and $\Ao_w$ of the
scalar field $\Phi(Q,t)$ will then be given by the same equation as 
(\ref{MM-scat0}),
\bea
\Ai_w = e^{iw\ln 4} \Ao_w + O(\gs)~.
\label{MM-scat}
\eea
\section{Bogolubov Analysis}
\label{s:bogolubov}
For our explicit analysis we shall consider the draining Fermi sea
background in which case we have in mind all the equations of the 
previous section with $\lp =\l,~ \lm=0$. 
On the matrix model side the wall scattering of the massless
particles can be computed \cite{polchinski91} very easily by exploiting
the fact that the
classical trajectories of the fermions on the phase-space are exactly
known. 
\begin{figure}
\begin{center}
\leavevmode
\hbox{%
\epsfxsize=3.0in
\epsfysize=3.0in
\epsffile{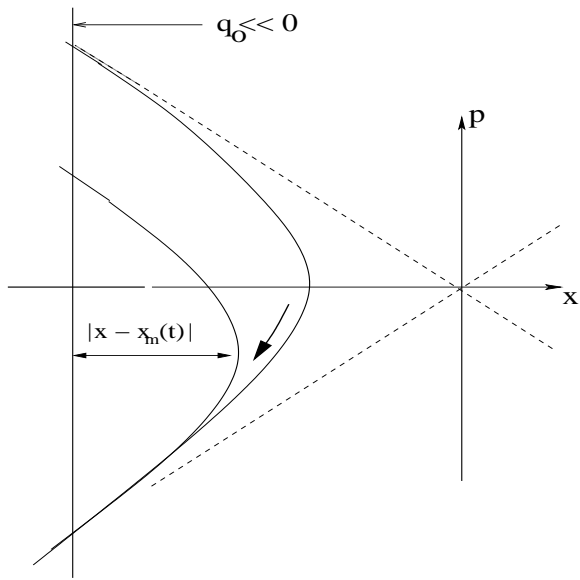}}
\caption{In the draining Fermi sea background the observer initially
sitting far away from the wall at $q_o <<0$ gradually enters the 
strongly coupled region.}
\label{fig-PP-DFS}
\end{center}
\end{figure}
Given an incoming pulse created on the upper edge of the Fermi 
sea at $q<<0$ one can compute the deformed outgoing pulse that
re-emerges, after some time on the lower edge 
without worrying about the complicated self interactions that the pulse 
goes through. In the draining Fermi sea background, as time evolves
the wall moves toward left (fig.\ref{fig-PP-DFS}) so that the observer
gradually enters the
strongly coupled region. We want to study the system over a time
span during which the observer remains in the free region. If a pulse
is sent in by the observer at a time $t_- (<<0)$ which is taken to be
of the same order of the observer's position $q_o(<<0)$ then the time
at which the pulse is received back is given by,
\bea
t_+ = -q_o - \ln(1+2\l)~. 
\label{t+}
\eea
The boundary of the space $q_m(t)$, which is zero at $t=t_-$, is
given, at $t=t_+$, by,
\bea
q_m(t_+) = -\ln 2\l - t_+ + O(1/\l e^{t_+}) = q_o - 
\ln \lt( {2\l \over 1 + 2\l} \rt)~,
\label{qmt+}
\eea
where we have taken $e^{q_o} << {\l \over 1 + 2\l}$. 
To keep $q_o$ in the free region at $t=t_+$ we should have
$q_o-q_m(t_+)<<0$. These two conditions can be written as,
\bea
-q_o >> \ln1/2\l >> 0~.
\label{condition} 
\eea
We shall see that to compute the Bogolubov coefficients we need to
work with sufficiently small $\l$ so as to be consistent with the
above equation. Then given a small value of $\l$ it also gives 
an estimation for the position of our observer. 
\subsection{Bogolubov Coefficients - Scattering Method}
\label{ss:scattering}
Let us now proceed to compute the Bogolubov coefficients. These are
the coefficients of the linear expansion of the out-oscillators
in terms of the in-oscillators where a mixing between the
positive and negative frequency modes occur. One may, therefore,
wonder if the leading order scattering equation obtained by the method
of \cite{polchinski91} can give these coefficients. In spite of the fact
that the analysis of \cite{polchinski91} gives only the tree-level 
results, with an additional input of coordinate transformation it is
indeed able to capture particle production which is a quantum effect.   
At late time, the natural oscillators in the draining Fermi sea
background are $\ao_w$, whereas at early time these are given by,
\bea
\ai_w = \Ai_w = e^{iw\ln 4} \Ao_w + O(\gs)~
\label{DFS-scat}
\eea
where the first equality is due to the fact that at early time the 
draining
background merges with the static one and in the second step we have
used the scattering equation (\ref{MM-scat}) in the static background.
The Bogolubov coefficients, therefore, effectively relate $\ao$ and
$\Ao$ oscillators.
\bea
\ao_w = \int_0^{\infty} d \wp \lt[\a (w, \wp) \Ao_{\wp} + \b (w, \wp)
\Aod_{\wp} \rt].
\label{def-bogol-I}
\eea
These coefficients can then in turn be computed by 
using the transformation,
\bea
\del_v \phi = \del_v U \del_U \Phi + \del_v V \del_V \Phi~,
\label{field-transf}
\eea
and setting 
$\del_U \Phi =0$ at late time which is the main use of
Seiberg bound in the present analysis.
For small $\l$ we get the following results (see appendix
\ref{A:bogolubov-scattering} for details),
\bea
\a(w, \wp) &=& \sqrt{w \over \wp} 
\int_{-\infty}^{\ln 1/2\l} {dx\over 2\pi} e^{iw x - 
i\wp f_{\l}(x)}~,\cr
\b (w,\wp) &=& \sqrt{w \over \wp} 
\int_{-\infty}^{\ln 1/2 \l} {dx\over 2\pi} e^{iw x + 
i\wp f_{\l}(x)}~,
\label{bogol-I}
\eea
where the function $\fl(x)$ is given by,
\bea
\fl(x) = x - \ln(1-2\l e^x)~, \quad -\infty < x <\ln1/2\l~.
\label{f}
\eea
\begin{figure}
\begin{center}
\leavevmode
\hbox{%
\epsfxsize=2.25in
\epsfysize=2.5in
\epsffile{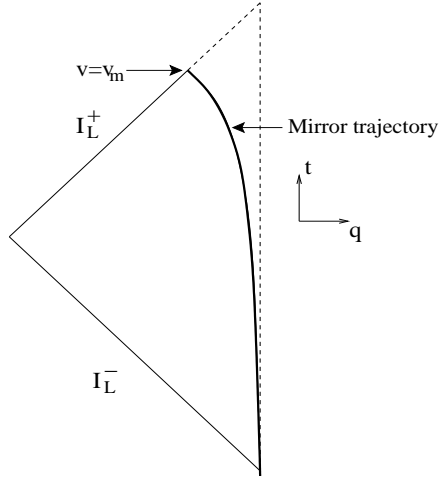}}
\caption{The asymptotic null infinities are given by, 
${\cal I}_L^- :(-\infty < u< \infty,~v=-\infty)$, 
${\cal I}_L^+:(u= \infty,~ - \infty <v \leq v_m =1/\ln2\l)$ and the 
mirror trajectory is given by, $u=\fl(v)$. The mirror approaches the
null line $v=v_m=\ln 1/2\l$ at late time. Taking $\l \to 0$ pushes
$v_m$ to infinity resulting in a trivial trajectory.}
\label{fig-mov-mirr}
\end{center}
\end{figure}
Several comments are in order.
\begin{enumerate}
\item
The results (\ref{bogol-I}) resemble the standard expressions for the
Bogolubov coefficients in a moving mirror problem in flat-space with
mirror-trajectory (fig.\ref{fig-mov-mirr}),
\bea
u = \fl(v)~.
\label{fake-traj}
\eea
Notice that this is not the actual trajectory given by $q_m(t)$. We
shall understand the physical origin of the function $\fl(x)$ in the
above equation when we analyze the moving mirror method in the next
section.
\item
Notice that,
\bea
\lim_{\l \to 0} \a(w, \wp) = \delta (w-\wp) ~, \quad \lim_{\l \to 0} 
\b(w, \wp) = 0 ~.
\label{stat-lim-ab} 
\eea
indicating no mixing of modes, hence no particle production. This
corresponds to the static background as the trajectory trivializes in
the above limit.
\item
We should emphasize that the above Bogolubov coefficients are
approximately correct for small but non-zero $\l$. The small $\l$ limit is
necessary for the following approximation which has been applied 
in our derivation (appendix \ref{A:bogolubov-scattering}) to invert a
certain equation. 
\bea
I_{\l}(w) &=& \int_{-\infty}^{\ln1/2\l} {dx\over 2\pi} ~ e^{iwx}~, \cr
&\sim & \delta (w) ~, \quad \hbox{for small } \l~.
\label{Il-approx}
\eea
\item
Explicit evaluation of the integrals in (\ref{bogol-I}) give,
\bea
\a(w,\wp) &=& {1\over 2} \sqrt{w\over \wp} \delta (w-\wp) +
{1\over 2\pi} e^{i(w-\wp)\ln 1/2\l}  
\sqrt{\wp \over w} {\Gamma(i(w-\wp)) \Gamma(i\wp) \over 
\Gamma(iw)} ~, \cr && \cr
\b(w,\wp) &=& {1\over 2} \sqrt{w\over \wp} \delta (w+\wp) 
-{1\over 2\pi} e^{i(w+\wp)\ln 1/2\l}  
\sqrt{\wp \over w} {\Gamma(i(w+\wp)) \Gamma(-i\wp) \over \Gamma(iw)}
~. \cr && 
\label{bogol-II}
\eea
\begin{itemize}
\item
Keeping in mind that the frequencies are always positive so that we
can set $\delta (w+\wp)=0$ one recovers the trivial result 
(\ref{stat-lim-ab}) in the limit $\l \to 0$. 
\item
For small but non-zero $\l$ the second term in each of the 
above expressions are highly fluctuating. The non-trivial feature that
the $\l$ dependence appears only in a phase makes it possible that certain
physical quantities in which this phase is canceled have $\l$
independent results (we shall see an example below). This is
consistent 
with the fact that $\l$ can
actually be changed by time translation and therefore a typical
time-averaged quantity should not be sensitive to this parameter.
Local quantities
which have to be given at a particular time do depend on
$\l$. Although it is not clear how to make sense of these
fluctuating objects in that case we argue in appendix \ref{A:relation} 
that the energy-momentum analysis give smooth $\l$ dependent functions
which can be related to some integrals of these Bogolubov coefficients.
\end{itemize}

\end{enumerate}

The above Bogolubov coefficients relate
the oscillators $\ao_w$ and $\Ao_w$. We may relate the ``string
theory'' oscillators $\abo_w$ and $\abi_w$ by 
introducing another set of coefficients $\a^S(w,\wp)$ and $\b^S(w,\wp)$,
\bea
\abo_w = \int_0^{\infty} d\wp \lt[\a^S(w,\wp) \abi_{\wp} + 
\b^S(w,\wp) \abid_{\wp}\rt].
\eea
Since at early time the present background approaches the static one
eq.(\ref{leg-pole}) may be used to relate the ``string theory'' and
``matrix model'' oscillators. But it is not clear if
eq.(\ref{leg-pole}) is still valid at late time. Assuming that this is
the case one can derive the following expressions,
\bea
\a^S(w,\wp) &=& {\cal P}(w,\wp) \a(w,\wp) ~, \cr
\b^S(w,\wp) &=& {\cal P}(w,-\wp) \b(w,\wp)~,
\eea
where ${\cal P}(w,\wp)$ is simply a pure phase,
\bea
{\cal P}(w,\wp) &=& e^{i(w-\wp)\ln 2} \frac{\Gamma(-iw)}{\Gamma(iw)}
\frac{\Gamma(-i\wp)}{\Gamma(i\wp)} ~,
\eea
so that the result for the total number of out-going particles between the 
frequency range $w$ and $w+dw$ remains the same,
\bea
N(w) &=& \int_0^{\infty} d \wp \lt| \b^S (w, \wp) \rt|^2 = 
\int_0^{\infty} d \wp \lt| \b (w, \wp) \rt|^2~,\cr
&=& \int_0^{\infty} {d\wp\over 4\pi (w+\wp)} {\sinh \pi w \over 
\sinh \pi \wp \sinh \pi (w+\wp)}~.
\eea
This is an example of a $\l$ independent physical quantity mentioned 
above. None of our expressions are infrared regularized. Apart from this
difficulty the above integral is well behaved. In fact, with an
infrared cutoff in the integral, $N(w)$ is finite for all $w$. 
\subsection{Bogolubov Coefficients - Moving Mirror Method}
\label{ss:moving-mirror}
We have seen in subsection \ref{ss:MM} that at the linearized level
the static background corresponds to a massless scalar on a
half-line with a flat metric. 
In this approximation the problem of particle production in 
the draining Fermi sea background reduces to that of a moving mirror 
problem in a non-flat metric \cite{karczmarek04, das04}. The metric 
$g_{\mu \nu}$ in $(t,q)$-coordinates can be obtained by transforming 
the metric, $G_{\mu \nu} = \hbox{diag(1,-1)}$, 
which corresponds to $(T,Q)$-coordinates
under the coordinate transformation (\ref{Q-q})
with $\lp=\l$ and $\lm=0$,
\bea
g_{\mu \nu} &=& {1\over (\cosh q-\l e^t)^2 -1} 
\lt( \begin{array}{cc}
\sinh^2 q - 2\l e^t \cosh q & \l e^t \sinh q \\ 
\l e^t \sinh q & - \sinh^2 q \end{array} \rt)
\label{metric-g0}
\eea 
At large time this reduces to,
\bea
g_{\mu \nu} &=& {1\over (1-2\l e^v)^2 - 4 e^{2q}} \lt(
\begin{array}{cc} 1 - 4\l e^v & -2\l e^v \\ 
-2\l e^v & -1 \end{array}
\rt)
\eea
Notice that at late time eqs.(\ref{qmt+}, \ref{condition}) give: 
$q_m(t_+) \approx q_o + \ln1/2\l << 0 $. Therefore for any 
$q < q_m(t)$ at which $(1-2\l e^v)$ is finite,
$e^{2q}$ is negligibly small. 
In this approximation the metric turns out to be,
\bea
g_{\mu \nu} &=& {1\over (1-2\l e^v)^2} \lt(
\begin{array}{cc} 1 - 4\l e^v & -2\l e^v \\ -2\l e^v & -1 \end{array}
\rt)~.
\label{metric-g}
\eea
Notice that even at finite time the metric (\ref{metric-g0}) reduces
to the above one at large $|q|$. This form precisely corresponds to
the coordinate transformation (\ref{UV-uv}) with $\lp = \l,~\lm=0$. 
At early time when $v\to -\infty$, $g_{\mu \nu} \to G_{\mu \nu}$. This
implies that the metric (\ref{metric-g}) which looks simpler than 
(\ref{metric-g0}) captures the correct coordinate systems in the
following region of validity: (1) everywhere at early time, (2) large
coordinate distances from the boundary at a finite time and (3)
coordinate distances from the
boundary such that $(1-2\l e^v)$ is finite at late time. 
To avoid complications we shall
replace the metric (\ref{metric-g0}) by (\ref{metric-g}) for all
space-time in our analysis. 
In this simplified model the mirror-trajectory is given by,
\bea
u=p(v)~, \quad p(v) = v-2\ln (1-2\l e^v)~.
\label{traj0}
\eea
Now the question is: given the metric (\ref{metric-g}) and the
mirror-trajectory (\ref{traj0}) what are the natural in and 
out-modes? In appendix \ref{A:modes} we have generalized the 
argument of geometrical optics for constructing these modes from 
flat-space to the present case. The results are,
\bea
\pin_w &=& {1\over \sqrt{4\pi w}} \lt[ e^{-iw \xi} + s~e^{-iw
  \fl(v)}\rt] \theta(q_m(t) - q) ~, \quad w\geq 0 ~,\cr
\po_w &=& {1\over \sqrt{4\pi w}} \lt[ e^{-iw \gl(\xi)} + s~
e^{-iw v}\rt] \theta(q_m(t)-q) ~, \quad w\geq 0~,
\label{pin-po-present}
\eea
where $\xi(u,v)$, given in eq.(\ref{null-rays1}), is one of the null 
coordinates in the present metric which replaces $u$ in the flat
case. The other null coordinate $v$ remains the same as in the flat
case. The functions $\fl(x)$ and $\gl(x)$, which are inverse of each
other $\fl(\gl(x)) =x $, are given in eqs.(\ref{f}) and (\ref{g}) 
respectively. Notice that, as discussed in appendix \ref{A:modes}
(eq.(\ref{traj-xi-zeta}) with $\zeta =v$ and $f(x) = \fl(x)$),
the function $\fl(x)$ characterizes the trajectory in terms of the null
coordinates $(\xi, v)$, which is the key reason for it to appear in the
expressions (\ref{bogol-I}) for the Bogolubov coefficients.

There remains the question of orthonormality of these
modes, a property which is needed to compute the Bogolubov coefficients.
We shall now discuss this issue. The ``time'' of our problem is
inherited from the matrix model to be the coordinate $t$. It can
be checked that this is a reasonable candidate for a ``time
function'' in the present case. Therefore the corresponding space-like 
surface $\Sigma$ 
is given by the constant $t$ surface coordinatized by $q$. Then the 
inner-product is 
given by\footnote{ Since $g = \det{g_{\mu \nu}} = g_{11}$,
the lapse function $N$ \cite{wald} is simply $1$ in our case.}
\bea
(\phi_1, \phi_2) &=& -i \int_{-\infty}^{q_m(t)} dq \sqrt{-g}~
g^{0 \mu} \phi_1 \del^{\leftrightarrow}_{\mu} \phi_2^* ~,\cr && \cr
&=& -i \int_{-\infty}^{q_m(t)} dq \lt[ \frac{1+2\l e^v}{1-2\l e^v} 
\phi_1 \del^{\leftrightarrow}_u \phi_2^* + \phi_1
\del^{\leftrightarrow}_v \phi_2^* \rt]~,
\label{inner-prod}
\eea
where we used the usual notation: $X \del^{\leftrightarrow} Y = 
X \del Y - (\del X) Y$. 
Using this inner product one can explicitly check that 
the in-modes (\ref{pin-po-present}) are always orthonormal, but the 
out-modes are orthonormal only for small $\l$ when
eq.(\ref{Il-approx}) is valid.
\bea
(\pin_{w_1}, \pin_{w_2}) = \delta (w_1-w_2)~, &\quad & 
(\pin_{w_1}, \pins_{w_2}) = 0~, \quad ~~~~~\hbox{always}~, \cr
(\po_{w_1}, \po_{w_2}) = \delta (w_1-w_2)~, &\quad & 
(\po_{w_1}, \pos_{w_2}) = 0~, \quad \hbox{for small $\l$ .} 
\label{ortho-in}
\eea
In the computations of $(\po_{w_1}, \po_{w_2})$ and 
$(\po_{w_1}, \pos_{w_2})$ the function $I_{\l}(w)$ (see
eq.(\ref{Il-approx})), with proper arguments, appears at various places. In the
small $\l$ limit, which is relevant for our discussion 
(see eq.(\ref{condition})), one uses eq.(\ref{Il-approx}) to achieve 
orthonormality\footnote{The key reason behind the lack of
orthonormality for a finite $\l$ is that $\Ip_L$ is not a full line, 
rather it is truncated due to the presence of the horizon at 
$v=v_m=\ln 1/2\l$ (fig.\ref{fig-light-cone}). Notice that for the
cases where the mirror reaches a constant velocity less than
that of light the whole portion of $\Ip_L$ is available and $\po_w$ 
constructed in appendix \ref{A:modes} are orthonormal. 
This problem also exists in the usual moving-mirror analysis in flat
space \cite{davies77}.}. Recall that the computation of the Bogolubov 
coefficients in the scattering method, described in appendix
\ref{A:bogolubov-scattering}, also went through only for small $\l$.

In the present method one can compute the 
Bogolubov coefficients by computing certain inner-products of the 
in and out-modes. Defining
\bea
\a(w, \wp) \equiv (\pin_{\wp}, \po_w)~, \quad \b(w, \wp) \equiv 
(\pins_{\wp}, \po_w)~,
\label{def-bogol-modes}
\eea
we get the same results as reported in eqs.(\ref{bogol-I}).
\section{Stress-Tensor Analysis}
\label{s:stress-tensor}
So far we have studied the problem through the Bogolubov method. 
By construction this method addresses the question of how many particles
are created in a time dependent background. A field theory also comes
with a natural definition of an energy-momentum tensor. In a time
dependent background one can ask how the vacuum expectation value of
the energy and momentum densities evolve with time. This is the
question that we shall be interested in. 
\subsection{Energy and Momentum Densities in Point-Splitting
Regularization}
\label{ss:densities}
The unregularized energy-momentum tensor is given by, 
\bea
T_{\mu \nu} &=& \del_{\mu} \phi \del_{\nu} \phi - 
{1\over 2} g_{\mu \nu} T ~,\cr
T &=& g^{\mu \nu} \del_{\mu} \phi \del_{\nu} \phi~,
\label{Tmunu}
\label{em}
\eea
The canonical Hamiltonian and the total momentum are defined to 
be\footnote{The symmetrization between the indices $\mu$ and $\nu$ in
the third equation does not have any effect on the finite part of the 
vacuum expectation value. Certain linearly divergent terms 
(\ref{TmunuS}) obtained in the point-splitting regularization get 
canceled because of this symmetrization.},
\bea
H&=& \int_{-\infty}^{q_m(t)} dq \sqrt{-g} ~\P_{0}~,\quad 
P= \int_{-\infty}^{q_m(t)} dq \sqrt{-g} ~\P_{1}~,\cr
\P_{\mu} &=& {1\over 2} ~g^{0\nu} (T_{\nu \mu} + T_{\mu \nu})~,
\label{Pmu}
\eea
Working in the Heisenberg picture we always take the vacuum to be the 
in-vacuum,  the one that is annihilated by $\ai_w$'s corresponding to 
the modes $\pin_w$ given in eqs.(\ref{pin-po-present}). The vacuum
expectation values of the above operators can simply be obtained by
mode expanding the scalar field. We use the in-modes and adopt a
symmetrical point-splitting regularization to separate out the
singular and the regular parts. 
\bea
\quad a_1 = a-{\delta a\over 2} ~, \quad a_2 = a+ {\delta a \over 2}~,
\quad a = u, v~. 
\label{point-split}
\eea
The details of this procedure are
given in appendix \ref{A:regularization}. 
Below we quote the final results for the energy and momentum densities.
The region of validity of these results is same as that mentioned
below eq.(\ref{metric-g}).
\bea
\E &\equiv & \sqrt{-g} ~\L \P_0\R = \E_S + \E_R~, \cr
\E_S &=& - {1\over 4\pi (\dv)^2} \lt[1+ {1-\hl^2 \over (\gamma
-\hl)^2}\rt]~, \cr
\E_F &=& - {\psi(\hl) (1-\hl^2) \over 2 \pi (\gamma -\hl)^3} + 
{(\hlp)^2 \over 16 \pi (\gamma-\hl)^2} - {S(\fl ,v)\over 24 \pi}~.
\label{E}
\eea
\bea
\P &\equiv & \sqrt{-g} ~\L \P_1\R = \P_S + \P_R~, \cr
\P_S &=& -{1\over 4\pi (\dv)^2} \lt[1 - \lt(1+\hl \over \gamma - \hl
\rt)^2 \rt]~, \cr
\P_F &=& {\psi(\hl) (1+\hl)^2 \over 2\pi (\gamma -\hl)^3}
+ {(\hlp)^2 \over 16 \pi (\gamma-\hl)^2} - {S(\fl,v) \over 24 \pi}~,
\label{P}
\eea
where the subscripts $S$ and $F$ refer to the singular and the finite
parts. The point-splitting is given by $\delta v$ and the limit
$\delta v \to 0$ is understood. The path along which the points are
split is given by the function $\gamma(u,v)$.
\bea
\delta u = \gamma(u,v) \delta v ~.
\label{path}
\eea
Using the metric (\ref{metric-g}) one can check that the path is null 
whenever $\gamma = \hl$, which is also expected from
the results (\ref{E}, \ref{P}) \cite{birrell}. 
The functions $\hl(v)$, $\psi(\hl)$ and the Schwarzian $S(f,x)$
are given by,
\bea
\hl(v) &=& -\del_v \xi = 2 \l e^v \fl^{\prime}(v)~, \cr
\psi(\hl)&=& {\hl \over 24} + {\hl^2 \over 8} + {\hl^3 \over 12}~, \cr
S(f,x)&=&  {f^{\prime \prime \prime}(x) \over f^{\prime}(x)} - 
{3\over 2} \lt( {f^{\prime \prime}(x) \over f^{\prime}(x)}\rt)^2~,
\label{h-psi-S}
\eea
where the primes denote derivatives with respect to the
argument. Notice that since we have used the in-modes (which are
always orthonormal), formally the results (\ref{E}, \ref{P}) do not 
need a small $\l$ approximation. But as explained at the beginning of
sec.\ref{s:bogolubov} the linearized approximation to the equation of
motion may not be good at a finite $\l$.
\subsection{Collective Field Theory Regularization in Static
Background}
\label{ss:static-reg}
Physical conclusion can not be drawn from the results (\ref{E},\ref{P})
as we have not yet fixed the subtraction scheme. Since we simply have a
scalar field in a background metric, normally this procedure will be
ambiguous. But as we mentioned toward the end of subsection
\ref{ss:MM} that being a dual description of a string theory, the 
collective field theory Hamiltonian automatically comes with a
particular counter term (\ref{Hct}) which fixes the subtraction scheme. 
Below we shall find a way to incorporate this subtraction scheme in the
present method of computation through mode expansion. This is easier
to do in the static background which we discuss first turning to the
draining background in the next subsection.  

Since we are dealing with a free ($\gs \to 0$) theory the
renormalization is only at the one loop level. Therefore 
we simply replace $\del_x \bar \phi(x,t)$ in the counter term
(\ref{Hct}) by the background solution (\ref{coll-backgrd}).
\bea
\Delta H = {1 \over 2 \pi} \int_{-\infty}^{-1} dx P_0(x)
\del_x \del_{x^{\prime}} \ln |x-x^{\prime}|_{x=x^{\prime}}~,
\label{Hct1}
\eea 
where the function $P_0(x)$ is defined in eq.(\ref{p0pm}).
Moreover, the vacuum expectation value of the renormalized energy
density (in $q$ space) is given by 
\cite{das90, gross912}\footnote{At large negative $q$,
where the effective coupling is small and the present analysis is
reliable, one gets, $\E^{Coll}_R \sim -1/24 \pi$. },
\bea
\E^{Coll}_F &=& - {P_0(x)^2 \over 12 \pi} S(q,x) ~, 
\label{coll-T00} 
\eea
where the function $q(x)$ is given in eq.(\ref{x-q}). 
At early time one gets from (\ref{E}),
\bea
\E_S = - {1\over 4 \pi (\delta v)^2} \lt(1+{1\over \gamma^2}\rt)~,
\quad
\E_F = 0~.
\eea
Now consider the following point-splitting regularization,
\bea
\E^{Coll} = {P_0(x)^2 \over P_0(x_1) P_0(x_2)} \E ~,\quad 
\gamma(u,v) = -1 ~,
\label{static-reg}
\eea
Noticing that $P_0(x)^2 = 1/(q^{\prime}(x))^2$ we write,
\bea
\E^{Coll} = - P_0(x)^2  
{q^{\prime}(x_1) q^{\prime}(x_2) \over 2 \pi (q(x_1) - q(x_2))^2}~.
\eea
Then using the following identity,
\bea
{ K^{\prime}(x-\delta x/2) K^{\prime}(x+\delta x/2) \over 
\lt(K(x-\delta x/2) - K(x+\delta x/2) \rt)^2} 
&=& {1\over (\delta x)^2 } + {1\over 6} S(K, x) 
+ {\cal  O}(\delta x)~, 
\label{identity}
\eea
for a function $K(x)$ we arrive at,
\bea
\E^{Coll} = -{P_0(x)^2 \over 2\pi (\delta x)^2 } - {P_0(x)^2
\over 12\pi} S(q, x)~.
\label{static-T00}
\eea
The first term gets canceled precisely by the counter term in the
collective field theory Hamiltonian (\ref{Hct1}) so that the 
renormalized energy density is same as that given in eq.(\ref{coll-T00}).
\subsection{Collective Field Theory Regularization in Draining
Background}
\label{ss:draining-reg-coll}
\begin{figure}
\begin{center}
\leavevmode
\hbox{%
\epsfxsize=1.9in
\epsfysize=2.5in
\epsffile{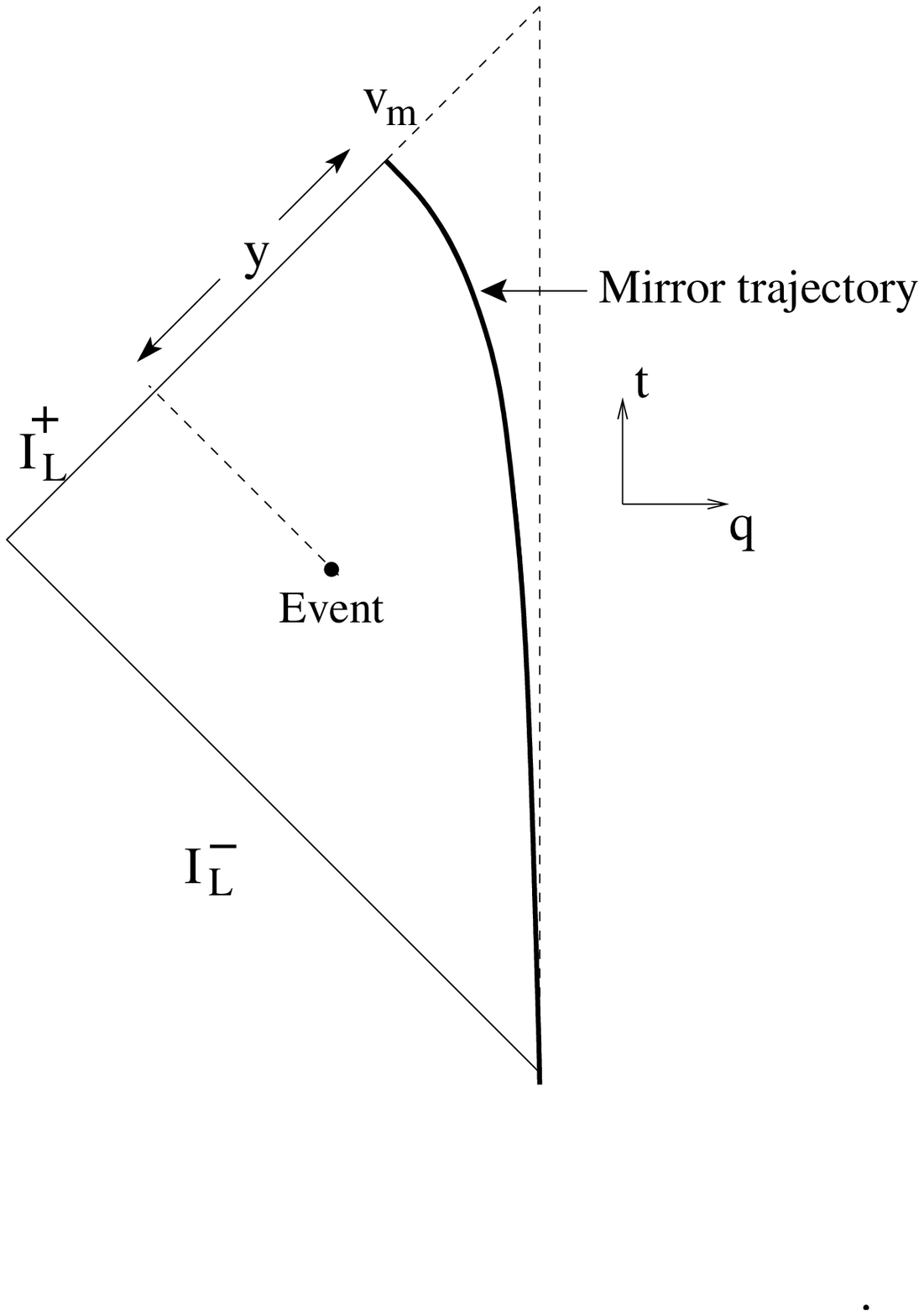}}
\caption{$y=|v-\ln1/2\l|$ gives the coordinate distance between the
projection of an event on ${\cal I}_L^+$ and the boundary on 
${\cal I}_L^+$. At late time, since 
$q_m(t) \approx - t + \ln1/2\l$, $y$ is same as the coordinate distance
from the boundary.}
\label{fig-y}
\end{center}
\end{figure}

Here we generalize the regularization (\ref{static-reg}) to the
draining Fermi sea background. 
Given the solution (\ref{coll-backgrd}) for the static background we
may write the regularization (\ref{static-reg}) in terms of this
solution which can then be naturally generalized to,
\bea
\E^{Coll} = \E_S^{Coll} + \E_F^{Coll} &=& 
{\lt( \del_x \bar \phi_0^{(D)}(x,t) \rt)^2 \over 
\del_{x_1} \bar \phi_0^{(D)}(x_1,t_1) 
~\del_{x_2} \bar \phi_0^{(D)}(x_2,t_2)} \E ~,\cr
&=& 
{P_0(X)^2 \over P_0(X_1) P_0(X_2)} \E_S + \E_R ~,
\label{draining-reg}
\eea
where we have used the solution for the draining Fermi sea,
\bea 
\del_x \bar \phi_0^{(D)}(x,t) = {P_0(X) \over \sqrt{\pi} \gs}~,
\label{coll-backgrd-D}
\eea
which can be obtained following the discussion in section \ref{s:MFS}
and noticing that $\del_x = \del_X$.
To compute the singular and the finite parts $\E_S^{Coll}$ and  
$\E_F^{Coll}$ respectively one uses the following 
results\footnote{Various coordinates like $X, Q$ can be recalled from
eqs.(\ref{coord-transf}, \ref{X-Q-x-q}, \ref{Q-q}, \ref{UV-uv}) with
$\lp=\l, \lm=0$. As argued below eq.(\ref{metric-g}) we can take
eq.(\ref{UV-uv}) to be correct for our simplified model. This gives the
second equation.},
\bea
{P_0(X)^2 \over P_0(X_1) P_0(X_2)} {1\over (\dv)^2} &=& 
P_0(X)^2 \lt[{1\over (\delta X)^2} + {1\over 6} S(Q,X) \rt]
\lt({\delta Q \over \dv} \rt)^2~, \cr
\lt( {\delta Q \over \dv} \rt)^2 &=& 
{1\over 4} (1-\gamma +2\hl)^2 + (1-\gamma + 2\hl) \psi(\hl) (\delta
v)^2 + O(\dv^4)~. \cr &&
\eea
Notice that $X_1$ and $X_2$ are not symmetrically separated from
$X$. Therefore using the identity (\ref{identity}) directly gives the
Schwarzian term evaluated at $\bar X$ which sits midway between $X_1$
and $X_2$. But since this term is not multiplied by any singular
term it is finally evaluated at $X$ in the coincidence limit.
The results for $\E_S^{Coll}$ and $\E_F^{Coll}$ turn out to be,
\bea
\E_S^{Coll} &=& - {P_0(X)^2\over 2\pi (\delta X)^2} \eta (\gamma, \hl)~, \cr
\E_F^{Coll} &=& - {P_0(X)^2 \over 12 \pi} S(Q,X) \eta(\gamma, \hl) 
- {\psi(\hl)(\gamma^2 -2\gamma \hl + 1) 
\over \pi (\gamma -\hl)^2 (1-\gamma + 2 \hl)} + \E_F~, \cr
\eta(\gamma, \hl) &=& {1\over 8} (1-\gamma+2\hl)^2 
\lt[1 + {1-\hl^2 \over (\gamma - \hl)^2 }  \rt]~. 
\label{ES-ER1}
\eea
A similar analysis for the momentum density gives the following
results,
\bea
\P_S^{Coll} &=& - {P_0^2(X) \over 2\pi (\delta X)^2} 
\bar \eta(\gamma,\hl)~, \cr
\P_F^{Coll} &=& - {P_0^2(X) \over 12 \pi} S(Q,X) \bar 
\eta(\gamma, \hl) 
- {\psi(\hl)(\gamma^2 -2\gamma \hl - 1) 
\over \pi (\gamma -\hl)^2 (1-\gamma + 2 \hl)}
+ \P_F ~, \cr
\bar \eta(\gamma, \hl) &=& {1\over 8}(1-\gamma +2\hl)^2 
\lt[ 1-\lt( {1+\hl \over \gamma-\hl }\rt)^2 \rt]~.
\label{Pcoll}
\eea
The Schwarzian terms in the above equations can be 
obtained by using the result,
\bea
P_0^2(X) S(Q,X) = {1\over 2} + {3\over 2 P_0^2(X)}~,
\eea
and noticing that in the region of validity of our analysis
so that we have,
\bea
\E_F^{Coll} &=& - {\eta(\gamma, \hl) \over 24 \pi} - 
{\psi(\hl)(\gamma^2 -2\gamma \hl + 1) 
\over \pi (\gamma -\hl)^2 (1-\gamma + 2 \hl)}+ \E_R~, \cr
\P_F^{Coll} &=& -{\bar \eta(\gamma,\hl) \over 24 \pi} - 
{\psi(\hl)(\gamma^2 -2\gamma \hl - 1) 
\over \pi (\gamma -\hl)^2 (1-\gamma + 2 \hl)}+ \P_R~.
\label{ER-PR-Coll}
\eea
Considering a spatial splitting ($\gamma(u,v) =-1$, which we shall justify
below (\ref{gamma})) we see that both the above results depend only on $v$. 
In fact defining,
\bea
y &\equiv & -(v - \ln1/2\l)~, \cr
\th(y) &\equiv &  \hl(v(y)) =  {e^{-y} \over 1- e^{-y}}~,
\label{def-y}
\eea 
the results in (\ref{ER-PR-Coll}) can be expressed entirely in terms
of $y$ (fig.\ref{fig-y}). 
The results turn out to be quite simple for $\gamma(u,v)=-1$,
\bea
\E_F^{Coll} = -{1\over 24 \pi} - {5 \over 48 \pi} \th (y) ~, \quad
\P_F^{Coll} = - {\th(y) \over 16 \pi} ~,
\label{ER-PR-Coll1}
\eea
so that at large $y$ both the above quantities fall off
exponentially to the values corresponding to the static background.

Notice that the final results (\ref{ER-PR-Coll1}) very much depend on
what regularization we work with. We shall now justify our
regularization (\ref{draining-reg}) by showing that in the region of
validity of our analysis, it is consistent
with the collective field theory counter term.
The original derivation \cite{andric85} of the collective field theory
counter term (\ref{Hct}) was formally background independent. This
would imply that the counter term for the draining Fermi sea
background will be given by,
\bea
\Delta \Hl &=& {\gs \over 2 \sqrt{\pi}} 
\int_{-\infty}^{x_m(t)} dx ~
\del_x \bar \phi_0^{(D)}(x,t)  ~
\del_x \del_{x^{\prime}} \ln |x-x^{\prime}|_{x=x^{\prime}}~, \cr
&=& 
 {1\over 2 \pi} \int_{-\infty}^{x_m(t)} dx ~P_0(x+\l e^t)~
\del_x \del_{x^{\prime}} \ln |x-x^{\prime}|_{x=x^{\prime}}~.
\label{Hlct} 
\eea
In the regularization (\ref{draining-reg}) we have an undetermined 
function  $\gamma(u,v)$. As long as we take 
$\lim_{v\to -\infty} \gamma(u,v) \to -1$ we recover both the singular and 
the finite parts of eq.(\ref{static-T00}) from eqs.(\ref{ES-ER1}). 
But for a Hamiltonian formulation it will be more natural to consider 
a spatial splitting. This requires us to take for all $u$ and $v$,
\bea
\gamma(u,v) = -1 ~.
\label{gamma}
\eea
With this one can compute $\E_S^{Coll}$ given in eqs.(\ref{ES-ER1}).
To cancel this particular value of $\E_S^{Coll}$ one needs the
following counter term, 
\bea
(\Delta \Hl)_{our} = {1\over 2 \pi}\int_{-\infty}^{x_m(t)} 
dx ~\lt\{{P_0(X)
(1+\hl) \over P_0(x)} \rt\} ~{P_0(X) \over (\delta x)^2}~.
\label{Hlct1}
\eea
It is straightforward to check that the factor in the curly bracket 
simply reduces to $1$ in the region of validity giving the same result 
as (\ref{Hlct}).
\section{Conclusion}
\label{s:conclude}
The problem of particle production in the draining Fermi sea
background of the two dimensional string theory was considered in 
\cite{das04} (see also \cite{karczmarek03}). In this paper the same
problem has been reconsidered and analyzed in more detail. The basic
accomplishments are the following:
\begin{enumerate}
\item
Physical arguments have been given to establish that the Bogolubov
analysis is sensible only in a limit where the deformation parameter
$\l$ is small but nonzero. 
\item
An additional method of computing Bogolubov coefficients, namely the
scattering method (which is very reminiscent of Polchinski's wall
scattering method \cite{polchinski91} in the static background), has 
been presented.
\item
The moving mirror method of computing the Bogolubov coefficients has
been elaborated through the explicit construction of the natural
out-modes. We show that the small $\l$ limit that has been 
considered on physical grounds is required to hold in order
for both the above methods (which are quite independent) to work.
\item
Time evolution of the energy-momentum tensor in the linearized
approximation of the equation of motion has also been studied in detail.
In \cite{das04} the coordinate transformation (\ref{UV-uv}) 
(with $\l_+= \l~,\l_-=0$) was taken to be,
\bea
U=u~, \quad V = v- \ln (1-2\l e^v)~,
\label{conf}
\eea
which is a conformal transformation. This, in fact, seems to be a good
approximation\footnote{I wish to thank S. R. Das for mentioning
another point of view where eq.(\ref{conf}) stands for the exact 
coordinate transformation. In that case one studies physics in a completely
different coordinate system whose ``time'' is different from the
original matrix model ``time''. It is not very clear what the
consequences of this approach would be.} on ${\cal I}_L^+$ 
(where $u \to \infty$) 
at a finite value of $\ln(1-2\l e^v)$. The present work attempts to 
go beyond this
approximation by incorporating the effect of the second term in the 
expression for $U$ in eq.(\ref{UV-uv}). This term takes the
transformation off the conformality giving a non-trivial effect on the
metric (see eq.(\ref{metric-g0})). 

The present computations have been done with a metric (\ref{metric-g})
which is simpler than the actual one. This procedure gives rise to a 
reduced region of validity
given by, (1) everywhere at early time, (2) large $|q|$ at a finite
time and (3) finite $(1-2\l e^v)$ at late time. 
Our computation is done in a particular regularization which leads to 
the certain singular term that is required to cancel the collective 
field theory counter term for the present background. As expected, 
this cancellation takes place only in the region of validity mentioned
above.
 
The final results show that the energy and momentum densities have
non-trivial time evolution. They start up with constant values at
early time but grows exponentially at late time as one approaches the
boundary from a large distance. This clearly indicates radiation of 
energy due to the space-time decay. This is different from the 
result found in \cite{das04}\footnote{Our result includes the
contribution considered in \cite{das04} in the form of a Schwarzian 
term (see, for example, the 
last term in the second equation of \ref{Fab}). In addition to this it
receives non-trivial contributions due to the existence of a non-flat 
metric and the particular subtraction scheme considered.}
where the densities remain constant as time evolves. 

Similar remarks apply to the result found in 
\cite{karczmarek04, karczmarek042}. The only difference between the 
treatments of \cite{karczmarek04} and \cite{das04} is that in 
\cite{karczmarek04} the initial value of $T_{vv}$ (which corresponds
to the static background) was taken to be zero as opposed to the case 
of \cite{das04}. As a result \cite{karczmarek04} also finds
exponential fall-off of the energy density. Although this is similar
to the result found in this paper, the reasons are different.

\end{enumerate}

\bigskip

{\bf Acknowledgment}:
I am thankful to S. R. Das for many discussions and various useful
comments on the manuscript. I wish to thank J. Michelson and
A. Shapere for enthusiastic questions and comments during a seminar 
delivered at the Department of Physics and Astronomy, University of 
Kentucky where the results of this paper were reported. I also thank 
A. Sen for his useful comments on the manuscript. This work was
supported by National Science Foundation grant PHY-0244811 and the 
Department of Energy grant No. DE-FG01-00ER45832.

\appendix
\sectiono{Bogolubov Coefficients in Scattering Method}
\label{A:bogolubov-scattering}
Here we give the details of the computation of the Bogolubov
coefficients defined in eq.(\ref{def-bogol-I}). Using 
eq.(\ref{field-transf}) with $\del_U \Phi =0$, the coordinate 
transformation 
(\ref{UV-uv}) with $\lp =\l,~ \lm=0$ and the mode expansion for the 
field we get,
\bea
\int_0^{\infty} dw \sqrt{w} \lt[\Ao_w e^{-iwV} - h.c. \rt]
&=& \int_0^{\infty} dw \sqrt{w} \lt[\ao_w 
{e^{-iwg_{\l}(V)} \over 1+ 2\l  e^V} - h.c. \rt]~,
\label{Ai-ao}
\eea
where we have introduced the function,
\bea
g_{\l}(x) = x - \ln (1 + 2 \l e^x) ~, \quad -\infty < x < \infty~.
\label{g}
\eea
Using the expansion (\ref{def-bogol-I}) in eq.(\ref{Ai-ao}) and 
performing an indefinite integral over $V$ on both sides one gets,
\bea
{1\over \sqrt{w}} e^{-iwV} = \int_0^{\infty} {d\wp \over \sqrt{\wp}}
\lt[\a (\wp , w) e^{-i\wp \gl(V)} + \b^*(\wp,w) e^{i\wp \gl(V)} \rt]
\label{eval-bogol-I}
\eea
Introducing the function $\fl(x)$, the inverse of $\gl(x)$, as 
in eq.(\ref{f}) we write the identity 
(\ref{eval-bogol-I}) as,
\bea
 {1\over \sqrt{w}} e^{-iw f_{\l}(x)} =
\int_0^{\infty} {d\wp \over \sqrt{\wp}} \lt[\a (\wp, w) e^{-i\wp x}
+ \b^*(\wp , w) e^{i\wp x} \rt] ~.
\label{eval-bogol-II}
\eea
Multiplying both sides by $e^{i\wb x}$ and integrating over the
valid range of $x$ (see eq.(\ref{f})) one gets,
\bea
{1\over \sqrt{w}} 
\int_{-\infty}^{\ln 1/2 \l} {dx \over 2\pi} e^{i\wb x - iwf_{\l}(x)}
&=&
\int_0^{\infty} {d\wp \over \sqrt{\wp}} \lt[\a(\wp ,w) 
I_{\l}(\wb -\wp) + \rt. \cr &&
\lt. ~~~~~~~~~~~~\b^*(\wp , w) I_{\l}(\wb + \wp) \rt]~, 
\label{eval-bogol-III}
\eea
where,
\bea
\Il(w) = \int_{-\infty}^{\ln1/2\l} {dx \over 2\pi} e^{iw x}~,
\label{Il}
\eea
which, in the small $\l$ limit, approaches $\delta (w)$. Approximating
$\Il(w)$ by $\delta (w)$ we get the coefficient $\a(\wb, w)$ from 
eq.(\ref{eval-bogol-III}). Similarly multiplying both sides of 
eq.(\ref{eval-bogol-II}) by $e^{-i\wb x}$ and following the same steps
one evaluates the coefficient $\b(\wb, w)$. 
The results that one gets are given in eqs.(\ref{bogol-I}).
\sectiono{Geometrical Construction of the In and Out-Modes}
\label{A:modes}
In a moving mirror problem in flat space the natural in and out-modes 
have the
following geometrical optics interpretation. Each of the two kinds of modes 
is a superposition of an incident and a reflected waves. The incident
wave in an in-mode takes the simple plane-wave structure on
${\cal I}^-_{L}$, but the reflected one is complicated on ${\cal I}^+_L$.
For an out-mode the incident part is complicated on ${\cal I}^-_{L}$
so that after reflection it takes the simple plane-wave form
on ${\cal I}^+_L$. This means that if the metric (\ref{metric-g}) 
were flat the 
trajectory (\ref{traj0}) would have corresponded to the following
natural modes,
\bea
\pin_w &=& {1\over \sqrt{4\pi w}} \lt[ e^{-iw u} + s~e^{-iw
  p(v)}\rt] \theta(q_m(t) - q) ~, \quad w\geq 0 ~,\cr
\po_w &=& {1\over \sqrt{4\pi w}} \lt[ e^{-iw p^{-1}(u)} + s~
e^{-iwv}\rt] \theta(q_m(t)-q) ~, \quad w\geq 0~,
\label{pin-po-flat}
\eea
where $p^{-1}(x)$ is the inverse function of $p(x)$ and $s=\mp$
correspond to the Dirichlet and Neumann boundary conditions respectively. 
We now try to generalize the above construction to our case through
the following steps,
\begin{enumerate}
\item
Find the two null directions $R_{\pm}^{\mu}(u,v)$ which can be
found, up-to overall factors, by solving the following equations,
\bea
R_{\pm}^{\mu}(u,v)R_{\pm \mu}(u,v) = 0~.
\label{null-directions}
\eea
We get,
\bea
R_+^{\mu}\sim (1~, ~~1-4 \l e^v) ~, \quad R_-^{\mu} \sim ( 1~, ~~-1)~.
\label{Rpm}
\eea
These give the structure of the light-cones (fig.\ref{fig-light-cone})
in the given space. 
\begin{figure}[!h]
\begin{center}
\leavevmode
\hbox{%
\epsfxsize=3.0in
\epsfysize=3.0in
\epsffile{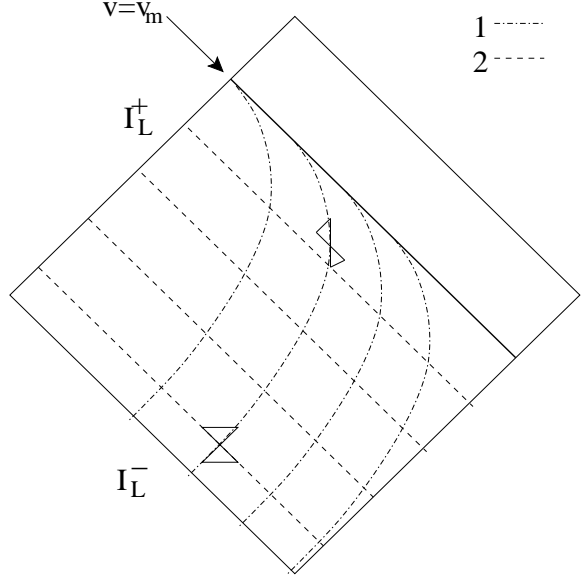}}
\caption{Curves with dot-style 1 and 2 are the null-rays whose 
tangent vectors are given by $R_+^{\mu}$ and $R_-^{\mu}$ respectively. 
Light-cones are same as that of a flat space on ${\cal I}^-_L$. 
They shrink as $v$ increases and are completely squeezed at
$v = v_m=\ln1/2\l$ where $R_+^{\mu} = R_-^{\mu}$.}
\label{fig-light-cone}
\end{center}
\end{figure}
Since the metric becomes flat as $v\to -\infty$, the 
light-cones are stretched to $90^{\circ}$ cone-angle in that region.
They shrink as $v$ increases and become completely squeezed at
$v=v_m=\ln 1/2\l$ where a horizon is formed\footnote{Notice that this is
true only for the simplified metric in (\ref{metric-g}) and not the
actual one given in eq.(\ref{metric-g0}). For the actual metric the 
light-cones are completely squeezed at the trajectory of the mirror
itself where the metric is singular.}. 
\item
Find the null rays 
\bea
\xi (u,v) = \hbox{const.}\quad \zeta(u,v) = \hbox{const.}~,
\label{null-rays}
\eea
i.e. the the curves which have $R_{\pm}^{\mu}(u,v)$ as tangent vectors
respectively.
\bea
R^{\mu}_+ \del_{\mu} \xi(u,v) = 0~, \quad R^{\mu}_- 
\del_{\mu} \zeta(u,v) = 0~.
\label{null-tangents}
\eea
The two null coordinates are found to be,
\bea
\xi(u,v)= u + \ln (1-2\l e^v) ~, \quad \zeta(u,v) = v 
\label{null-rays1}
\eea
\item
Generalize eqs.(\ref{pin-po-flat}) by the following,
\bea
\pin_w &=& {1\over \sqrt{4\pi w}} \lt[ e^{-iw \xi} + s~e^{-iw
  f(\zeta)}\rt] \theta(q_m(t) - q) ~, \quad w\geq 0 ~,\cr
\po_w &=& {1\over \sqrt{4\pi w}} \lt[ e^{-iw g(\xi)} + s~
e^{-iw\zeta}\rt] \theta(q_m(t)-q) ~, \quad w\geq 0~,
\label{pin-po-curved}
\eea
where the functions $f(x)$ and $g(x)$ are inverse of each other. $f(x)$ 
appears in the equation of the trajectory in the $(\xi, \zeta)$
coordinates in the following way,
\bea
\xi = f(\zeta)~.
\label{traj-xi-zeta}
\eea
It is straightforward to check that the functions $e^{iw\xi(u,v)}$ and 
$e^{iw\zeta(u,v)}$ are the two independent solutions of the
Klein-Gordon
field equation in the given metric. In our case the functions $f(x)$
and $g(x)$ appearing in eqs.(\ref{pin-po-curved})
turn out to be $\fl(x)$ and $\gl(x)$ introduced in eqs.(\ref{f}) and
(\ref{g}) respectively so that the modes take the final form as shown 
in eqs.(\ref{pin-po-present}).
\end{enumerate}
\section{Point-splitting Regularization in the Draining Fermi 
Sea Background}
\label{A:regularization}
It will be useful for 
our purpose to work in the $(u,v)$ coordinate system. The metric in
this system is given by,
\bea
\g_{ab} = {1\over 2} 
\lt(\begin{array}{cc} 0 & \flp(v) \\ \flp(v) & - 4 \l e^v \flp(v)^2
\end{array} \rt)~, \quad \g^{ab} = 2
\lt(\begin{array}{cc} 4 \l e^v & {1\over \flp(v)} \\ {1\over \flp(v)}
& 0 \end{array} \rt)~, 
\label{g-uv}
\eea
where $a,b = u,v $. We get,
\bea
T = 8\l e^v \del_u \phi \del_u \phi + {2\over \flp(v)} \del_u \phi
\del_v \phi + {2\over \flp(v)} \del_v \phi \del_u \phi ~.
\label{T}
\eea
Here we are distinguishing $\del_u \phi \del_v \phi$ from 
$\del_v \phi \del_u \phi$ as we shall consider a point-splitting
regularization of the operators. The components of the stress-tensor
in the $(u,v)$ system turn out to be,
\bea
T_{uu} &=& \del_u \phi \del_u \phi~,\cr
T_{vv} &=& \del_v \phi \del_v \phi + 2 \hl(v)^2 \del_u \phi \del_u
\phi + \hl(v) \lt(\del_u \phi \del_v \phi + \del_v \phi \del_u \phi \rt)~,\cr
T_{uv} &=& {1\over 2} \lt(\del_u \phi \del_v \phi - \del_v \phi \del_u
\phi \rt)- \hl(v) \del_u \phi \del_u \phi ~, \cr
T_{vu} &=& -{1\over 2} (\del_u \phi \del_v \phi - \del_v \phi \del_u
\phi )- \hl(v) \del_u \phi \del_u \phi ~,
\label{Tuv}
\eea
where $\hl(v)$ has been defined in eqs.(\ref{h-psi-S}).
At early time ($v\to -\infty \Rightarrow \hl(v) \to 0$) we 
recover the standard results for flat space. The vacuum 
expectation values can be written in terms of the basic objects,
\bea
D_{ab}(u,v) \equiv \L \del_a \phi \del_b \phi \R ~.\label{Dab}
\eea
For explicit computation we expand $\phi$ in terms of the in-modes. 
Then we point-split $D_{ab}(u,v)$ in the following way,
\bea
D_{ab}(u,v) = \lim_{\du, \dv \to 0} 
\int_0^{\infty} dw~ \del_{a_1} \pin_w(u_1,v_1)
\del_{b_2} \pins_w(u_2, v_2)~, 
\label{splitDi}
\eea
where the point-splitting is taken to be symmetrical (\ref{point-split}).
Using the full in-modes given in eqs.(\ref{pin-po-present}) and the 
result $\int_0^{\infty} dw~ w e^{iwx} = -{1\over x^2}$ one gets,
\bea
D_{uu} &=& - {1\over 4 \pi (\delta \xi)^2} ~, \quad \delta \xi = \xi_2
  -\xi_1~, \cr
D_{vv} &=& - {\hl(v_1) \hl(v_2) \over 4 \pi (\delta \xi)^2} -
{\flp(v_1) \flp(v_2) \over 4\pi \lt(\fl(v_1) - \fl(v_2) \rt)^2}
+ {s\over 4 \pi} \lt[{\hl(v_1) \flp(v_2) \over (\fl(v_2) - \xi_1)^2} 
+ {\hl(v_2) \flp(v_1) \over (\fl(v_1)-\xi_2)^2} \rt]~,\cr
D_{uv} &=& {\hl(v_2) \over 4 \pi (\delta \xi)^2 } - {s\over 4\pi}
\lt[ {\flp(v_2) \over (\xi_1-\fl(v_2))^2}\rt] ~,\cr
D_{vu} &=& {\hl(v_1) \over 4 \pi (\delta \xi)^2 } - {s\over 4\pi}
\lt[ {\flp(v_1) \over (\fl(v_1)-\xi_2)^2}\rt]~.
\label{splitDi1}
\eea
Notice that the terms in the square brackets do not have any short
distance divergence. Moreover, because of the factor in the
denominator, they are suppressed both at late time 
and at a finite time but large $|q|$. 
Therefore we shall simply drop these terms. The rest of the terms are 
computed, using the identity (\ref{identity}), to be,
\bea
D_{ab} &=& S_{ab} + F_{ab}~,
\label{D-SF}
\eea
where the singular and finite terms $S_{ab}$ and $F_{ab}$ respectively
are given by,
\bea
S_{uu} &=& - {1\over 4 \pi (\gamma -\hl)^2 (\dv)^2} ~, \cr
S_{vv} &=&  - {\hl^2 \over 4 \pi (\gamma - \hl)^2 (\dv)^2} 
-{1 \over 4 \pi (\dv)^2}~, \cr
S_{uv} &=& {\hl \over 4 \pi (\gamma - \hl)^2 (\dv)^2} +
{\hlp \over 8 \pi (\gamma-\hl)^2 \dv}~, \cr
S_{vu} &=&{\hl \over 4 \pi (\gamma - \hl)^2 (\dv)^2} -
{\hlp \over 8 \pi (\gamma -\hl)^2 \dv}~,
\label{Sab}
\eea
\bea
F_{uu} &=& -{\psi(\hl) \over 2 \pi (\gamma -\hl)^3 }~,\cr
F_{vv} &=& - {\hl^2 \psi(\hl) \over 2\pi (\gamma - \hl)^3} - 
{\hl \hlpp - (\hlp)^2  \over 16 \pi (\gamma-\hl)^2} - 
{S(\fl, v) \over 24 \pi}~,\cr
F_{uv}&=& F_{vu}= {\hl \psi(\hl) \over 2 \pi (\gamma-\hl)^3} + 
{\hlpp \over 32 \pi (\gamma -\hl)^2 }~, 
\label{Fab}
\eea
where $\psi(\hl)$ is given in eqs.(\ref{h-psi-S}). $\hl$ appearing on 
the right hand sides of all the above
equations is meant to be evaluated at $v$. We have also used 
eq.(\ref{path}).
These results and the standard transformation laws of 
stress-tensor components under
the coordinate transformation $(u,v) \to (t,q)$ allow us to finally
compute $\L T_{\mu \nu}\R$. The results are,
\bea
\L T_{\mu \nu}\R = \L T_{\mu \nu}\R_S + \L T_{\mu \nu}\R_F~,
\label{vac-Tmunu}
\eea 
where the singular and finite parts $\L T_{\mu \nu}\R_S$ and 
$\L T_{\mu \nu}\R_F$ respectively are given by,
\bea
\L T_{00}\R_S &=& -{1\over 4 \pi (\dv)^2} \lt[ 1 + 
\lt({1-\hl \over \gamma -\hl} \rt)^2\rt]~, \cr
\L T_{11}\R_S &=&  -{1\over 4 \pi (\dv)^2} \lt[ 1 + 
\lt({1+ \hl \over \gamma - \hl} \rt)^2\rt]~, \cr
\L T_{01} \R_S &=&  -{1\over 4 \pi (\dv)^2} \lt[ 1 -
{1-\hl^2 \over (\gamma -\hl)^2 } \rt] + {\hlp \over 4 \pi \dv (\gamma
-\hl)^2}~, \cr
\L T_{10} \R_S &=&  -{1\over 4 \pi (\dv)^2} \lt[ 1 -
{1-\hl^2 \over (\gamma -\hl)^2 } \rt] - {\hlp \over 4 \pi \dv (\gamma
-\hl)^2}~,
\label{TmunuS}
\eea
\bea
\L T_{00} \R_F &=& - {\psi(\hl) (1-\hl)^2 \over 2 \pi (\gamma -\hl)^3} 
+ {(\hlp)^2 \over 16 \pi (\gamma-\hl)^2} - {S(\fl, v) \over 24 \pi}~, \cr
\L T_{11} \R_F &=& - {\psi(\hl) (1+\hl)^2 \over 2 \pi (\gamma -\hl)^3} 
+ {(\hlp)^2 \over 16 \pi (\gamma-\hl)^2} - {S(\fl, v) \over 24 \pi}~, \cr
\L T_{01} \R_F &=& \L T_{10} \R_F = {\psi(\hl) (1-\hl^2) \over 
2 \pi (\gamma -\hl)^3} + {(\hlp)^2 \over 16 \pi (\gamma-\hl)^2} - 
{S(\fl, v) \over 24 \pi}~.
\label{TmunuR}
\eea
Notice that each of $\L T_{01}\R_S$ and $\L T_{10} \R_S$ has a
linearly divergent term which comes from the similar terms in $S_{uv}$
and $S_{vu}$ given in eqs.(\ref{Sab}). 
\section{Relation between Bogolubov and Stress-Tensor Analysis}
\label{A:relation}
Some part of $\L T_{\mu \nu} \R_R$ can, in fact,
be written in terms of the Bogolubov coefficients $\a(w,\wp)$ and 
$\b(w,\wp)$. Following \cite{davies77} we expand the right hand side 
of eq.(\ref{Dab}) in terms of the out-modes and use the unitarity 
properties of the Bogolubov coefficients to get,
\bea
D_{ab} &=& \Do_{ab} + B_{ab} ~, \cr
B_{ab} &=& \Brho_{ab} +\Bmu_{ab} ~,
\label{D-Do-B}
\eea
where 
\bea
\Do_{ab} &=& \int_0^{\infty} dw ~ \del_a \po_w \del_b \pos_w ~, \cr
\Brho_{ab} &=& 2  \lt[\int_0^{\infty} dw_1 dw_2 ~\del_a \po_{w_1} 
\del_b \pos_{w_2} \rho_{\l} (w_1,w_2) \rt]~, \cr
\Bmu_{ab} &=& 2 \hbox{Re} \lt[ \int_0^{\infty} dw_1 dw_2 ~ \del_a \po_{w_1}
\del_b \pos_{w_2} \mu_{\l} (w_1, w_2) \rt]~, \cr
\rho_{\l}(w_1, w_2) &=& \int_0^{\infty} dw \b(w_1, w) \b^*(w_2,w)~, \cr
\mu_{\l} (w_1,w_2) &=& \int_0^{\infty} dw \b(w_1, w) \a(w_2,w)~.
\label{Do-B}
\eea
Proceeding in the same way as the one that leads to eqs.(\ref{D-SF})
one can compute $\Do_{ab}$. Subtracting this off from $D_{ab}$ given in 
eqs.(\ref{D-SF}) one gets $B_{ab}$. The answers that one gets are the
following,
\bea
B_{uu} &=& {1\over 24 \pi} S(\gl, \xi)=-{1\over 48 \pi}  ~, \cr
B_{vv} &=& {\hl^2 \over 24 \pi} S(\gl, \xi) - {1\over 24 \pi}
S(\fl,v) = -{\hl^2 \over 48 \pi} - {1\over 24 \pi} S(\fl,v) ~, \cr
B_{uv}&=& B_{vu} = - {\hl \over 24 \pi} S(\gl ,\xi) = {\hl \over 48 \pi}~,
\label{Bab}
\eea
where we have taken the large time limit ($\lim_{u \to \infty} S(\gl,
\xi) \to -1/2 $) in the above expressions. Notice that $B_{ab}$ can
also be computed using eqs.(\ref{D-Do-B}) and (\ref{Do-B}) in terms of
the Bogolubov coefficients. 
This requires some particular integrals of these coefficients to be
equal to functions appearing on the right hand side of
eq.(\ref{Bab}). It can be explicitly checked that all such
relations can be satisfied by the following single equality.
\bea
\Irho_{\l}(v) + \Imu_{\l}(v) &=& - {1\over 12} S(\fl, v)~,
\label{equality}
\eea
where,
\bea
\Irho_{\l}(v) &=& \int_0^{\infty} dw_1 dw_2 \sqrt{w_1 w_2} 
e^{i(w_2-w_1)v} \rho_{\l}(w_1,w_2) ~, \cr
\Imu_{\l}(v) &=& \hbox{Re} \lt[ \int_0^{\infty} dw_1 dw_2 \sqrt{w_1 w_2} 
e^{i(w_2-w_1)v} \mu_{\l}(w_1,w_2)\rt] ~.
\label{Irho-Imu}
\eea
For genuine Bogolubov coefficients satisfying the unitarity conditions 
the above relation must be true. Our Bogolubov coefficients given in 
eqs.(\ref{bogol-II}) are approximately correct for small
$\l$. Although we have not quite succeeded, we attempted to understand
how far the equality (\ref{equality}) can be satisfied by our
Bogolubov coefficients. Below we shall report on how far we have been
able to go.

Using the variable $y$ (\ref{def-y}) one can erase the appearance of
$\l$ on the right hand side of (\ref{equality}).
\bea
S(\fl ,v) = - \del_y \th(y) - {1\over 2} \th(y)^2~.
\label{S-th} 
\eea
This means that it should be possible to fully absorb the $\l$ and
$v$ dependence of $(\Irho_{\l}(v) + \Imu_{\l}(v))$ into $y$. Using 
the $\a$ and $\b$ coefficients in (\ref{bogol-II}) one can readily
check that it really happens for $\Irho_{\l}(v)$,
\bea
\Irho(y) &:=& \Irho_{\l}(v) = 
{1\over 4\pi} \int_0^{\infty} dw_1 dw_2~ {e^{i(w_1-w_2)y} 
\over \G(iw_1) \G(-iw_2)} \int_0^{\infty} dw {\G(i(w_1+w))
\G(-i(w_2+w)) \over \sinh \pi w}~, \cr &&
\eea
but not quite for $\Imu_{\l}(v)$,
\bea
\Imu_{\l}(v) &=& - {1\over 4\pi} \hbox{Re} \lt[\int_0^{\infty} dw_1
dw_2 e^{i(w_1-w_2)y}{\G(i(w_1+w_2)) \G(iw_2) \over \G(iw_1)} 
\lt\{w_2 e^{2iw_2\ln1/2\l} \rt\} \rt] \cr
&& -{1\over 4\pi} \hbox{Re} \lt[\int_0^{\infty} dw dw_1 dw_2 
e^{i(w_1-w_2)y }
{ \G(i(w_1+w)) \G(i(w_2+w))\over \G(iw_1) \G(1+iw_2)\sinh \pi w }i 
\lt\{w_2 e^{2iw_2\ln1/2\l} \rt\}  \rt]~,\cr&&
\eea
Notices that because of the factor kept in the curly brackets in the above
expression one can set, for small $\l$,
\bea
\Imu_{\l}(v) = 0~,
\eea
so that one is left with the identity,
\bea
\Irho(y) =  {1\over 12} \del_y \th(y) + {1\over 24} \th(y)^2~.
\eea
It has not been possible for us to check this identity either
analytically or numerically.

\end{document}